\documentclass[12pt,preprint]{aastex}

 \usepackage[]{graphicx}

 \shorttitle{Loss of star forming gas in SDSS galaxies}
 \shortauthors{Calura et al.}

\begin{document}


 \title{Evidence for progressive loss of star-forming gas in SDSS galaxies}


 \author{F. Calura$^1$, R. Jimenez$^{2,3}$, B. Panter$^{4}$, F. Matteucci$^{1,5}$, A. F. Heavens$^{4}$}
\affil{
$^1$ INAF - Osservatorio Astronomico di Trieste, via G.B. Tiepolo 11,  
34131 Trieste, Italy \\
$^2$Institute of Space Sciences (CSIC-IEEC)/ICREA, campus UAB, Bellaterra-08193, Spain \\
$^3$Dept. of Astrophysical Sciences, Peyton Hall, Princeton University, Princeton, NJ 08544, USA\\
$^4$SUPA, Institute for Astronomy, University of Edinburgh, Royal Observatory, Blackford Hill, Edinburgh EH9-3HJ, UK\\
$^5$Dipartimento di Astronomia-Universit\'a di Trieste, Via G. B. Tiepolo 11, 34131 Trieste, Italy
}
 \email{fcalura@oats.inaf.it}



\begin{abstract}
Using the star formation rates from the SDSS galaxy sample, extracted using the MOPED algorithm, and the empirical Kennicutt law relating star formation rate to gas density, we calculate the time evolution of the gas
fraction as a function of the present stellar mass.  We show how the gas-to-stars ratio varies with stellar mass, finding good agreement with previous results for smaller samples at the present epoch. For the first time we show clear evidence for progressive gas loss with cosmic epoch, especially in low-mass systems.  We find that galaxies with small stellar masses have lost almost all of their cold baryons over time, whereas the most massive galaxies have lost little.
Our results also show that the most massive galaxies have evolved faster and
turned most of their gas into stars at an early time, 
thus strongly
supporting a downsizing scenario for galaxy evolution.
\end{abstract} 



\section{Introduction}

In galaxies, the star
formation rates are known to depend on the the cold ($T < 200$ K) gas densities 
via the well-known empirical Kennicutt (1998, hereafter K98) relation.
According to this relation, the star formation rate surface density $\dot{\Sigma_{*}}$ is proportional
to a certain power of the gas surface density  $\Sigma_{gas}$,
according to the relation $\dot{\Sigma_{*}} \propto \Sigma_{gas}^{k}$,
with $k \sim 1.4$. This relation expresses the natural condition that the star formation rate depends on the amount of gas available; galaxies
dominated by young stellar populations have large reservoirs of gas, which can be turned into stars.
On the other hand, gas-poor
galaxies have consumed their gas in the
past and/or have ejected it from its cold state, perhaps into the intergalactic medium, and at the present time they do not exhibit
significant star formation.  In dense environments,  gas can also be stripped, owing to interactions with 
the intergalactic medium or to  encounters with other systems.\\ 
A fundamental quantity related to the gas accretion history of galaxies is the gas to stellar fraction (G/S).
An important study of this quantity is the one of Kannappan (2004, hereinafter K04), who determined G/S for
$\sim 35000$ galaxies from the Sloan Digital Sky Survey (SDSS) and 2MASS databases on the basis of photometric techniques.
A similar study on
a smaller sample of low mass galaxies has been performed by Geha et al. (2006, hereinafter G06). 
Previous important studies of the evolution of the star formation and gas content in galaxies 
at low and high redshift are the works by Erb et al. (2006a, 2006b) and Erb (2008). \\
By means of analytic chemical evolution models, Dalcanton (2007)  studied the 
chemical properties of local galaxies and put some quantitative constraints on their 
amount of gas infall and outflow. However, 
none of these studies was able to address how infall and outflow 
vary as a function of the galactic mass. \\
In the present paper, we start from the SDSS galaxy sample and by means
of the K98 relation and other simple physical assumptions, 
we aim at computing how the gas accretion history
of local galaxies depend on their stellar masses. 
The sample we are considering includes
$\sim 310000$ galaxies, larger than the one considered by K04 by a factor of $\sim 9$.
The strength of our method is that it is based on very few
free parameters, whose range of values can be tested very easily, allowing us to draw
robust  conclusions on some fundamental physical properties of galaxies and on their evolution with
cosmic time. We study how the gas mass of these galaxies has
varied throughout their evolution and compare our predictions with observational data available
in the literature. We focus on the present-day gas fractions and study how these quantities depend
on the galactic stellar mass, investigating how the `downsizing' character of the galaxy populations affects
their gas accretion history.\\
The paper is organized as follows:
in section 2, we present the galaxy sample and the MOPED algorithm used for the analysis.
In section 3, we describe the methods used to derive the gas masses for the galaxies of the SDSS catalogue.
In section 4, we presents our results and in section 5 we draw some conclusions.
We assume a concordance cosmology with $\Omega_{m}=0.24$,
 $\Omega_{\Lambda}=0.76$ (Spergel et al. 2007) and $h=0.71$.

\section{SDSS DR3 analysis with MOPED}
\label{SDSS_descr}
Following Panter et al. (2007a), we select galaxies from the third data release
(DR3) main galaxy catalogue of the SDSS.
In order to remove any bias and
simplify our $V_{\rm max}$ criterion (where $V_{\rm max}$ is the
maximum volume of the survey in which the galaxy could be observed
in the SDSS sample), we have cut our sample at
$\mu_r<23.0$. 
The 
surface brightness cutoff of 23 $mag/arcsec^{2}$ excludes gas-rich, low surface brightness galaxies. 
However, very gas-rich galaxies are present in our sample, as we will see later in Sect. 4.1. \\
At low redshifts, a small number of Sloan galaxies are
subject to shredding - where a nearby large galaxy is split by the photometric pipeline  into several smaller sources. \\
Shredding affects a negligible number of galaxies for $0.01<z<0.25$, for $z<0.01$ it could be as many as 10\% of galaxies. 
(SDSS collaboration, private communication). We have chosen to make a conservative cut at $z=0.02$ to avoid this problem.  
This also removes the problems of non Hubble-flow
peculiar velocities giving erroneous distances based on redshift,
which can have a significant effect on recovered stellar mass and
metal return.The total number of galaxies that satisfy our cuts is $299571$.\\
In order to estimate the completeness 
of the survey we
have used the ratio of target galaxies to those that have observed
redshifts (P. Norberg, Priv. Comm.). This does not allow for
galaxies which are too close for the targetting algorithm, and we
estimate this fraction at a $6\%$ level from the discussion in
Strauss et al. (2002). As a result of both these cuts, our effective
sky coverage is $2947$ square degrees.
We also removed from our analysis a set of wavelengths which may be
affected by emission lines, not modelled by the stellar population codes.\\
We use the $3$ \AA\, single stellar population models of Bruzual \& Charlot (2003)
as the basis for this study, but we emphasize that the star formation histories allowed are very flexible.  We assume an Initial Mass
Function (IMF) given by Chabrier (2003), which is very
successful at reproducing current observations in our galaxy.
We chose to apply a uniform velocity dispersion of 170 kms$^{-1}$ to
the 3\AA\ models, reflecting a typical value for the Main Galaxy
Sample. Although a  velocity dspersion of  170 km/s is too
large to represent the smallest mass galaxies in the sample,  
this discrepancy does not affect any of our results.\\
Physical parameters of galaxies have been derived from their observed spectra by means of the MOPED algorithm
(Heavens, Jimenez \& Lahav 2000; Reichardt, Jimenez \& Heavens 2001).
As with previous MOPED studies, for this analysis we have used a
single foreground dust screen.
We chose to use an LMC extinction law for the main analysis as given in
Gordon et al. (2003).

\subsection{Recovering star formation and metallicity history using MOPED}
\label{MOPED}
In the past, the SFH of galaxies was typically modelled by an
exponential decay with a single parameter - for more complex models
one or two bursts of formation were allowed. In fact, it would be
better not to put any such constraints on star formation,
particularly considering that each galaxy may have (as a result of
mergers) several distinctly different aged populations. Star
formation takes place in giant molecular clouds, which have a
lifetime of around $10^{7}$ years. Splitting the history of the
Universe into the lifetimes of these clouds give a natural unit of
time for star formation analysis, but unfortunately it would require
several thousands of these units to map the age of galaxies formed 13
billion years ago, and the lack of sensitivity of the final
spectrum to the detailed history would make any estimate of star
formation history extremely degenerate.  We choose a compromise
solution, where we allow 11 time bins in which the star formation
rate (SFR) can vary independently. This is still a difficult
computational problem, but the MOPED (Heavens et al. 2000) algorithm reduces
processing time to approximately 1 minute per spectrum on a standard
workstation. The boundaries between the 11 different bins used are
determined by considering bursts of star formation at the beginning
and end of each period (at a fixed metallicity) and set the
boundaries such that the fractional difference in the final spectrum
is the same for each bin.  This leads to a set of timebins $t_{i}$ 
which are
almost equally spaced in log(lookback time).  Nine timebins are spaced
with a ratio of log(lookback time) of 2.07 in this application of
MOPED, plus a pair of high-redshift bins to improve resolution at
$z>1$. The central ages for the timebins $t_{i}$ are 0.0139,
0.0288, 0.0596, 0.123, 0.256, 0.529, 1.01, 2.27, 4.70, 8.50 and 12.0
Gyr. 
Single-value decomposition analysis of SDSS spectra, 
as suggested by Ocvirk et al (2006), indicates
that only 2-5 components can be reliably recovered, not 11, but MOPED in
fact recovers $\le 5$ significant components in 98\% of cases (Tojeiro et al
2007).\\
The gas which forms stars in each time bin is also allowed to
have a metallicity which can vary independently. Extensive testing
(Panter et al. 2003, Panter et al. 2007) shows that for large samples the average star
formation history is recovered with good accuracy.  The Bruzual \& Charlot (2003)
models allow metallicities between
$0.02<\left(Z/Z_\odot\right)<1.5$. In order to investigate
metallicity evolution (Panter et al. 2007) no regularization or
other constraint is applied to the metallicity of the populations -
each different age can have whatever metallicity fits best. A
further complexity to the parametrization to deal with post-merger
galaxies which contain gas which has followed dramatically different
enrichment processes would be to have several populations with the
same age but independent metallicities.   It is possible to consider
a more complex parametrization, but again one risks degeneracies in
solution.  With 11 ages, 11 metallicities and the dust parameter,
the model has 23 parameters. The 23 dimensional likelihood surface
is explored by a Markov-Chain-Monte-Carlo technique outlines in Panter et al. (2003). 
Further information on the MOPED algorithm is contained in Panter (2005).\\
We take account of the 3 arcsec fibre aperture by correcting the spectrum upwards by
the ratio of the $r$ band Petrosian flux in the photometric survey to the $r$ band flux 
received by the fibre. Although for individual galaxies this is likely to fail, for the population 
as a whole there is evidence from the colours (Glazebrook et al. 2003) that there is no 
overall bias between the sampling of stellar populations by the spectroscopy and photometry.

\section{Derivation of the gas masses}
The MOPED algorithm allows us to determine the star formation history of each galaxy,
i.e. the star formation rate values at the end of the 11 timebins described above.
For  each timebin, we aim at determining the cold gas mass of any galaxy on the basis of its
SFR, by inverting the K98 relation, which links the
gas surface density to the SFR per unit area.
A similar technique was used by  Erb et al. (2006a) and Erb (2008) to derive the gas fractions for  a sample of star forming 
galaxies at $z\sim 2$ and to study the implications of the mass-metallicity relation observed at high 
redshift on the galactic gas accretion history, respectively. \\
Following K98, for any galaxy and at a given time,
the gas surface density $\Sigma_{gas}$, expressed in $M_{\odot} \, pc^{-2}$, depends on the SFR surface
density $\dot{\Sigma_*}$, expressed in $M_{\odot} \, yr^{-1} \, kpc^{-2}$ according to:

\begin{equation}
 \Sigma_{gas} = \big( \frac{ \dot{\Sigma_{*}} }{ 2.5 \times 10^{-4}}\big)^{0.714} \, \, \, \, \,    M_{\odot} \, pc^{-2}
\end{equation}

The MOPED catalogue provides the deprojected SFRs in units of $M_{\odot}/yr$, hence
to derive the corresponding surface densities
we need an estimate of the galactic scale-length radius as a function of the baryonic
mass. 
We calculate this quantity by means of the scaling relations by Mo, Mao \& White (1998) and we follow its evolution 
with lookback time. 
A major  difference between the works by Erb et al. (2006a) and ours concern mainly our 
detailed study of the scaling radius as a function of mass and redshift. \\
For each galaxy, we know the present day stellar mass $M_*$, whereas the scale length is likely to be set principally by the total mass.
We assume that each galaxy is embedded inside a dark matter halo of mass
$M = M_*/f$,
where $f$ is a parameter which we take to be a constant.  The value of $f$ and its dependence on mass and time are, in a sense, 
part of what we are trying to obtain, but here we use $f$ only to set the scale-length of the galaxy.  
As we will see later, the assumption for $f$ has a minor impact on our results. \\
Initially, we will make the assumption (relaxed later) that the baryons collapse into a disk, which is assumed  to be thin and in centrifugal balance, with
an exponential surface density profile $\Sigma_{tot}(R)= \Sigma_0 \exp(-R/R_{d})$.
$\Sigma_0 = \Sigma_{gas} + \Sigma_{*}$ is the central density of the disk, having
an angular momentum which is a fraction $f$ of the angular momentum of the dark matter halo. 

For the halo, we assume a Navarro, Frenk \& White (1997) mass profile and we take into account the effects
of the self-gravity of the disk. 
For each halo of mass $M$, the virial radius $R_{200}$ is defined as
\begin{equation}
R_{200}\equiv \frac{V_c}{10 H(z)}
\label{6}
\end{equation}
where
\begin{equation}
V_c \equiv \left[10 \, G \, M\, H(z)\right]^{1/3}
\end{equation}
where $G$ is the gravitational constant and $H(z)$ is the Hubble parameter as a function of the redshift $z$,
given by
\begin{equation}
H(z)=H_0[\Omega_{\Lambda}+(1-\Omega_{\Lambda}-\Omega_m)(1+z)^2
+\Omega_m(1+z)^3]^{1/2}
\label{2}
\end{equation}
where $\Omega_m,\Omega_{\Lambda}$ and $H_0$ are the matter density parameter,  the vacuum energy
parameter and the Hubble constant at $z=0$.\\

The scaling radius $R_d$ can be calculated as (Mo et al. 1998):
\begin{equation}
R_d=\frac{\lambda R_{200}f_{c_{vir}}^{-1/2}f_{R}(\lambda,c_{vir},f_{b})}{\sqrt{2}}.
\label{6}
\end{equation}
$\lambda$ is the spin parameter of the halo and depends on the total energy of the halo $E$,
its angular momentum $J$ and its mass $M$ according to:
\begin{equation}
\lambda=J|E|^{1/2}G^{-1}M^{-5/2}.
\label{8}
\end{equation}
The quantity $\lambda$ is likely to assume values in the range $0.01\le \lambda\le 0.1$ (Heavens \& Peacock (1988), Barnes \& Efstathiou (1987), Jimenez et al. 1998).
In this paper, we assume a typical value of $\lambda=0.05$.  Scatter in $\lambda$ would propagate as scatter in the gas fractions, but we average over many galaxies to obtain the average gas fractions.\\
$c_{vir}$ is the halo concentration factor, and is calculated following Bullock et al. (2001) and Somerville et al. (2006) i.e.
for each halo, by defining a collapse redshift $z_c$ as $M_{*}(z_c)=F \cdot M$.
$c_{vir}$  is given by $c_{vir}(M,z)=K(1+z_c)/(1+z)$, where $F$ and $K$ are two adjustable parameters.
Following Somerville et al. (2006), we assume $K=3.4$ and $F=0.01$.\\
To compute the quantities $f_{c_{vir}}$ and $f_{R}$, we use analytic fitting functions presented in Mo, Mao \& White (1998). \\
The use of these scaling relations described above is physically motivated and, as shown by Somerville et al. (2006), accounts
for the observed redshift evolution of the disk size-stellar mass relation from
the GEMS (Galaxy Evolution from Mophology and SED; Rix et al. 2004) survey out to $z\sim 1$. \\
For each galaxy, if $\psi$ is the SFR in units of $M_{\odot}/yr$ and if we assume for the
SFR surface density profile $\dot{\Sigma}_{*}(R)=\dot{\Sigma}_{*,0} \, \exp (-1.4 R/R_{d})$,
the central SFR surface density is then given by:
\begin{equation}
\dot{\Sigma}_{*,0} = \frac{\psi}{2 \pi (R_{d}/1.4)^{2}}.
\end{equation}
The gas surface density $\Sigma_{gas}$ is then determined by the K98 relation, and the gas mass $M_{gas}$
(in $M_{\odot}$) is given by:
\begin{equation}
 M_{gas} = \Sigma_{gas} \times 2 \pi  R_{d}^{2}.
\end{equation}
In section 4.1, we relax the assumption of an exponential disk profile, and show how the assumptions of a different surface density profile and of different scaling
relations affect the calculated galactic gas masses.\\
At this stage, we can see how the calculated gas mass $M_{gas}$ depends on the baryonic fraction $f$. 
The quantity $R_{d}/R_{200}$ as a function of $f$ has been calculated by Somerville et al. (2006) and 
can be fitted by an exponential function $R_{d}/R_{200} = 0.77 \exp(-4.7 f)$. 
The dependence of $M_{gas}$ on $f$  is then approximately 
\begin{equation}
M_{gas} \propto exp(-2.7 f)/f^{0.19}
\end{equation}
The dependence of the resulting gas masses on $f$ is hence very weak,  as a result of which 
any reasonable estimate for $f$ will suffice.\\
In setting the scale length, we take $f = 0.06$, in agreement with the recent results of Hoekstra et al. (2005),
based on the weak-lensing signals of isolated galaxies at low redshift.  This is also consistent with most of the cold baryons being in stellar form.
Several recent cosmic baryon budget calculations (Calura \& Matteucci 2004, Fukugita \& Peebles 2004, Shankar et al. 2006) support this
hypothesis, in particular for spiral disks and spheroids.

A potential difficulty of our study is that the analysis above assumes that the galaxy remains a single structure during its history,
whereas the star formation history deduced from the fossil record makes no statement about where the stars were
when they were formed. Hence a natural question to ask is how the possible presence of merging affects these results.
However, if we assume the scaling relation $R_d \propto M_*^{1/3}$ holds for any sub-units before merging,
the final gas-to-star ratio is almost independent of the amount of merging which has taken place.
To see this, let us assume that at some lookback time the observed galaxy was in several pieces $i=1,\ldots n$,
each containing a fraction $f_i$ of the final stellar mass.  If we assume the star formation rate is proportional to $f_i$,
then the gas mass in each sub-unit is proportional to $(\psi_i/R_{di}^2)^{0.714} R_{di}^2 \propto f_i^{0.9}$, so the total gas mass is modified to
\begin{equation}
M_{gas} \rightarrow M_{gas}\sum_{i=1}^n f_i^{0.9}.
\end{equation}
Since $\sum_i f_i = 1$, we see that for any reasonable merging history, the gas mass is virtually unchanged.
The assumptions we make here are certainly challengeable, but this calculation gives us confidence that the results are
likely to be robust to the merger history.\\

\section{Results}
\subsection{The evolution of the gas to stellar mass fractions}
\begin{figure*}
\includegraphics[height=15cm,width=15cm]{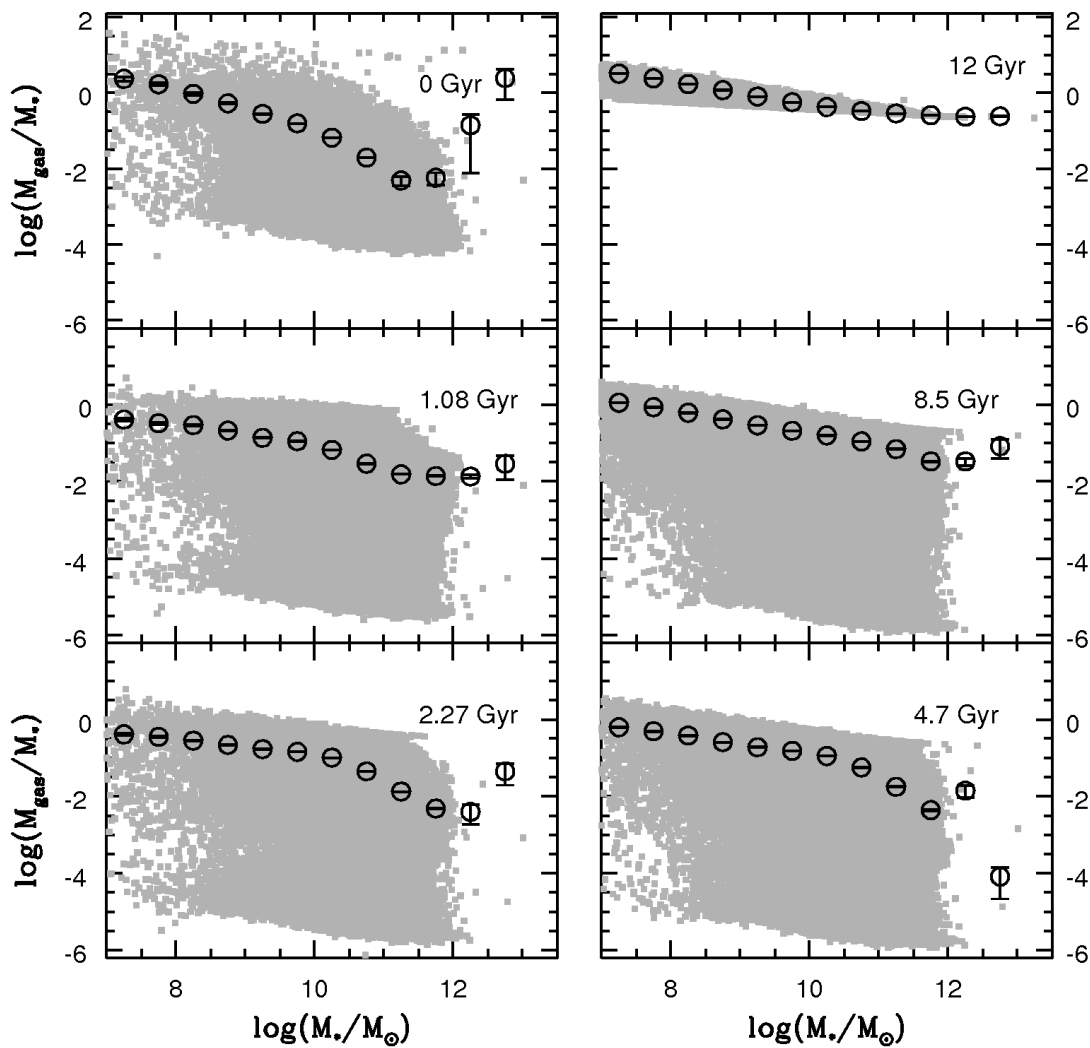}
\caption{Evolution of the gas to stellar mass ratios as
a function of the stellar mass
for the MOPED galaxies at various lookback times. The small grey squares represent the individual values,
whereas the black open circles with error bars represent the arithmetic means in various mass bins, plotted with their 
uncertainties. From the upper right panel,
in clockwise sense, we show the values computed at lookback times of 12 Gyr, 8.5 Gyr, 4.7 Gyr, 2.27 Gyr, 1.01 Gyr and 0 Gyr.  }
\label{plotone}
\end{figure*}

\begin{figure*}
\includegraphics[height=15cm,width=15cm]{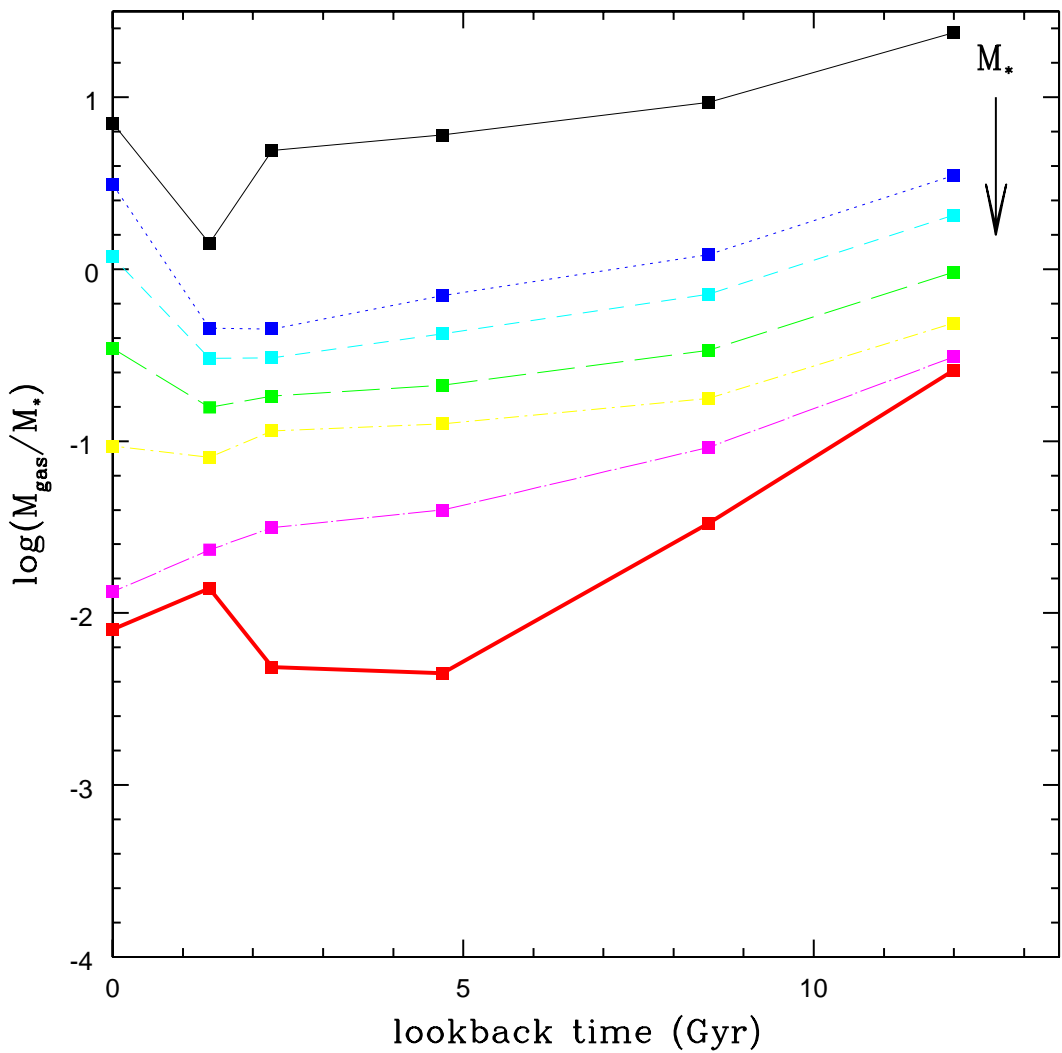}
\caption{The gas to stellar mass ratio as 
as a function of lookback time for 
several stellar mass $M_{*}$ bins.
Solid black thin line: $M_{*}\le 6.5 M_{\odot}$. 
Dotted blue thin line: $6.5 \le M_{*} /M_{\odot} \le 7.5 $. 
Short-dashed cyan thin line: $7.5 \le M_{*} /M_{\odot} \le 8.5 $. 
Long-dashed green thin line: $8.5 \le M_{*} /M_{\odot} \le 9.5 $. 
Dot-short-dashed yellow thin line: $9.5 \le M_{*} /M_{\odot} \le 10.5 $. 
Dot-long-dashed magenta thin line: $10.5 \le M_{*} /M_{\odot} \le 11.5 $. 
Solid black  thick line: $M_{*}  \ge 11.5 M_{\odot}$. }
\label{plottwo}
\end{figure*}

{ In this section, we present how the gas accrection history (i.e. the infall history) of our galaxies 
evolved with cosmic time. In section~\ref{ouflow}, we will present how the outflow history of our galaxies evolved with cosmic 
and with galactic mass. \\
One way to study the evolution of the gas accretion history is to analyze the evolution with mass and with cosmic time 
of the gas to stellar mass 
ratio (G/S).  }
In figure~\ref{plotone}, we show the evolution of G/S calculated as described in section 3, as
a function of the stellar mass
for the MOPED galaxies at various lookback times.
The dispersion of the calculated G/S is relatively small at
a lookback time of 12 Gyr. The
dispersion however remains large (the individual G/S span 6 orders of magnitude) up to the preset time.
At any lookback time, the average G/S ratio decreases
as the stellar mass increases in the range $10^{6} M_{\odot} \le M_{*} \le 10^{12} M_{\odot}$.
Note that the highest mass bins ($M_{*}> 10^{12} M_{\odot}$) are characterised by very few systems with a
peculiar behaviour. In the last Gyr of evolution, our predictions indicate an increase in
the average G/S ratios for the galaxies with stellar masses $M_{*}<10^{10} M_{\odot}$,
due to the fact that these galaxies must have accreted gas at late times.\\
The anti-correlation between (G/S) and stellar mass found at any lookback time can be interpreted in the 
following way. 
In general, higher mass galaxies are more efficient at turning gas into stars, and
lower mass galaxies are less efficient and so retain a large amount of
their primordial gas.  This is clearly visible also in Fig.~\ref{plottwo}, 
where we show the evolution of (G/S)  as a function of lookback time for
several mass bins.  In this figure, it is clear that 
the slope of the curves steepens from low mass to high mass systems, indicating decreasing gas consumption 
timescales from dwarf to giant galaxies. 
This behaviour is in agreement with the downsizing picture of galaxy evolution, according
to which the least massive systems formed the bulk of their stars (consuming most of their gas reservoirs)
at recent times, whereas in the most massive galaxies the buildup of the stellar mass was complete several
Gyrs ago.  \\
At the present time, we predict individual
$\log(G/S)$ values between $-4$ and $\sim 1.5$. This range of values is in agreement with existing
local observational estimates for galaxies of various morphological types.
The lowest estimates of G/S have been derived for local dwarf spheroidal galaxies and for large
ellipticals. The gas-poorest
dwarf spheroidal galaxies of the Local Group have upper limits on the atomic H to blue luminosity ratio
of $M_{HI}/L_{B} \sim 0.001$ (Mateo 1998). These values correspond to $\log(G/S) \sim -3.67$,
assuming a helium correction factor of 1.4 and a stellar mass-to-light ratio of
$(M/L)_{B,E} = 6.5 M_{\odot}/L_{\odot}$ (Fukugita, Hogan \& Peebles 1998). In local S0 and E galaxies, Sansom et al.
(2000) observed values of $M_{HI}/L_{B}$ down to $\sim 0.0001 M_{\odot}/L_{\odot}$, corresponding to
$\log(G/S) \sim -4.67$. \\
On the other hand, the largest gas reservoirs are observed in local
irregular galaxies. Hunter \& Elmegreen (2004) find for local irregulars $M_{HI}/L_{B}$
up to $\sim 5 M_{\odot}/L_{\odot}$, corresponding to $\log(G/S) \sim +0.8$,
assuming a helium correction factor of 1.4 and a stellar mass to light ratio of
$(M/L)_{B,irr} \sim 1 M_{\odot}/L_{\odot}$ (FHP98). 
These observational estimates are in very good  agreement with the lower and upper extremes of the
G/S values we derive for present-day galaxies.


\begin{figure*}
\includegraphics[height=15cm,width=15cm]{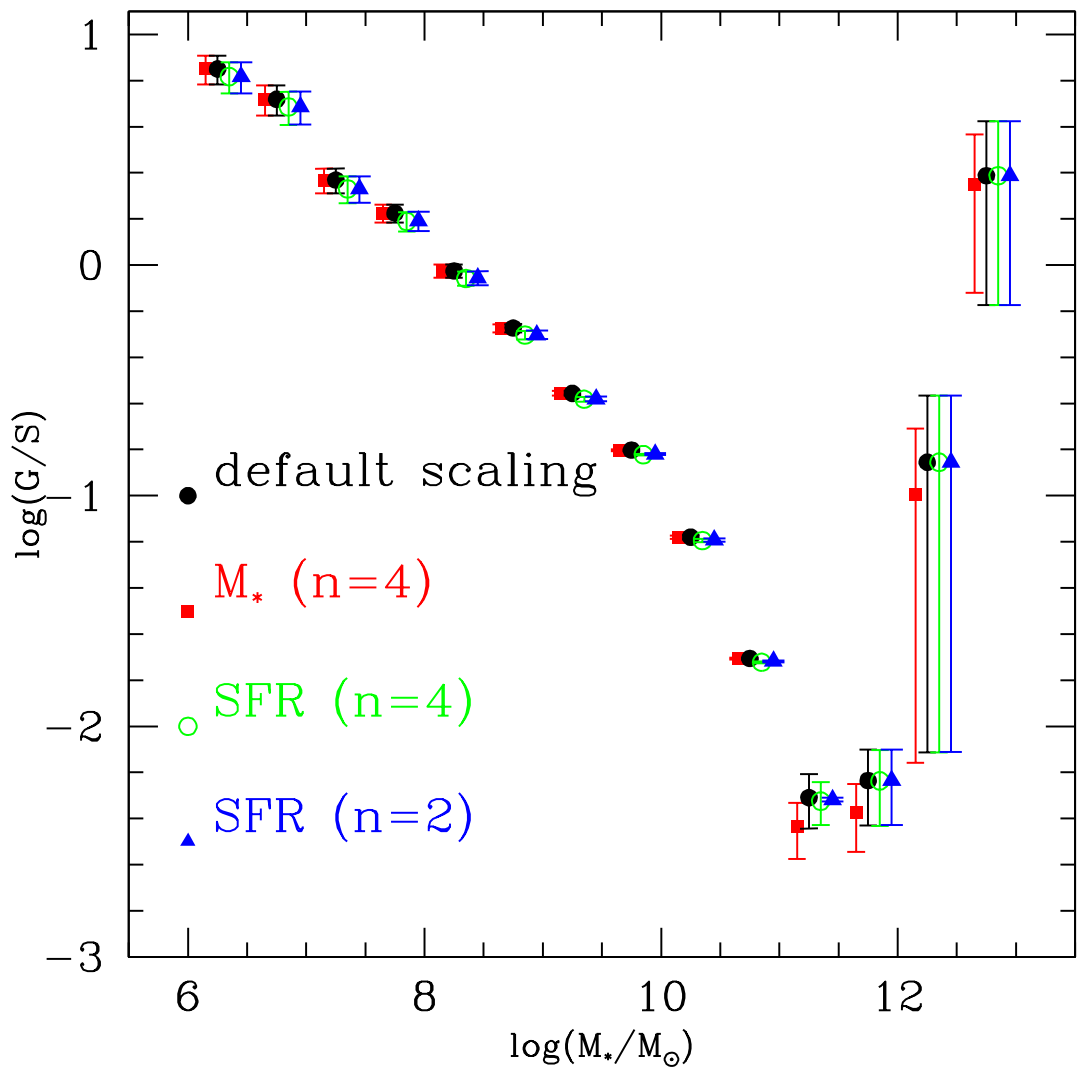}
\caption{The gas to stellar mass ratio as
a function of the stellar mass
for the MOPED galaxies, computed with the same default scaling relations for all galaxies
(solid circles), using different scaling relations for spheroids and spirals,
first disentangling by means of the stellar mass (solid squares, see text for details)
and secondly by means of the current SFR, assuming Sersic index $n=4$ (open circles) and  $n=2$ (solid triangles).}
\label{plotthree}
\end{figure*}
\begin{figure*}
\includegraphics[height=13pc,width=16pc]{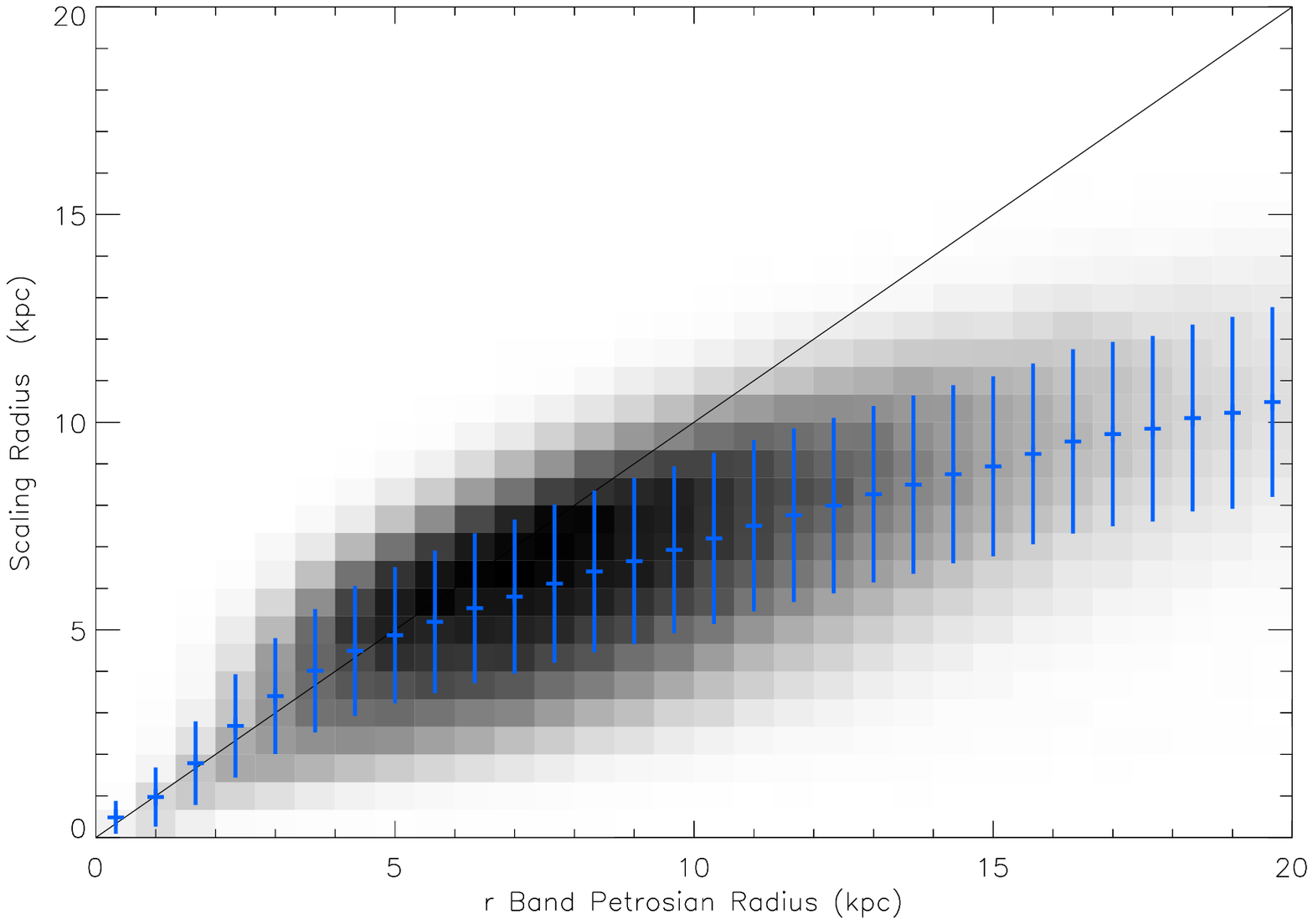}
\includegraphics[height=13pc,width=16pc]{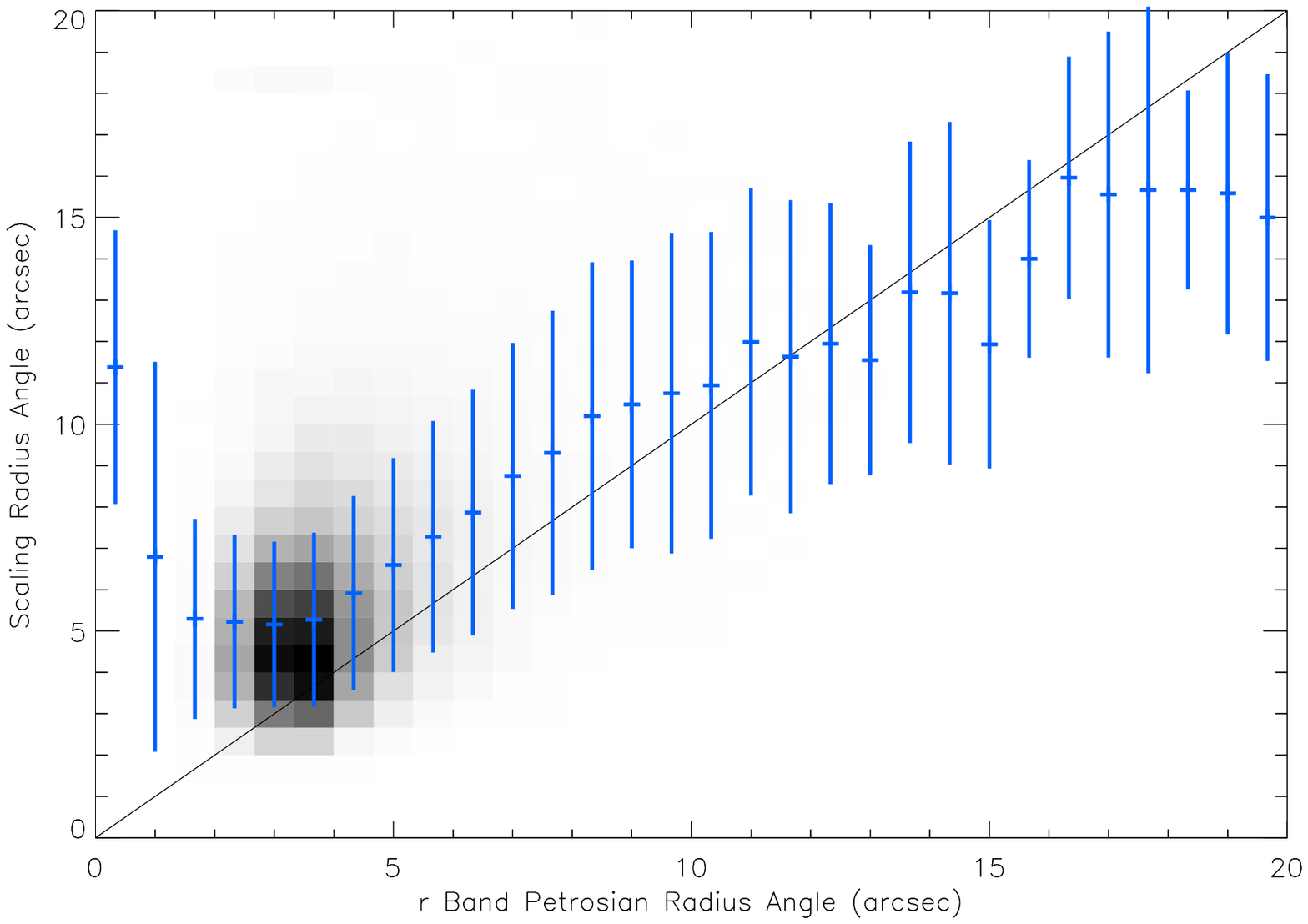}
\caption{\emph{Left panel}: Scaling radii as a function of the r-band Petrosian 
radii $R_P$ 
for  the SDSS sample, overplotted with the means $\pm$ 1 $\sigma$  errors. The darkest 
regions correspond to the densest areas, where most of our galaxies lie.  
\emph{Right panel}: expected angular size, computed  from the scaling radii, against actual angular size on the sky for the SDSS 
galaxies. }
\label{plotfour}
\end{figure*}
\begin{figure*}
\includegraphics[height=20pc,width=20pc]{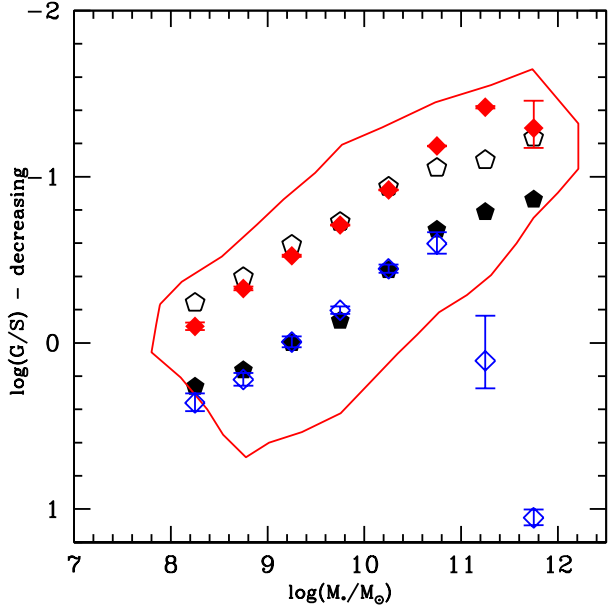}
\caption{G/S  as a  function of the stellar mass for the blue (solid pentagons)
and red (open pentagons) galactic components as computed by Kannapan (2004).
The contour encloses the area where 92\% of the
galaxies with stellar masses $9\le \log(M_{*}/M_{\odot})\le 12$ fall, after weighting each one by $1/V_{\rm max}$, according to K04.
The typical systematic error on the observational points is of 0.05 dex. The solid and open diamonds are the mean values computed
in this paper for the red and blue galactic components, respectively. In this case, the red and blue populations consist
of all the galaxies having formed more than the $60 \%$ and less than the $60 \%$ of
their present-day stellar mass during the first 4 timesteps (i.e. in $\sim 11.4$ Gyr), respectively.}
\label{plotfive}
\end{figure*}
\begin{figure*}
\includegraphics[height=20pc,width=20pc]{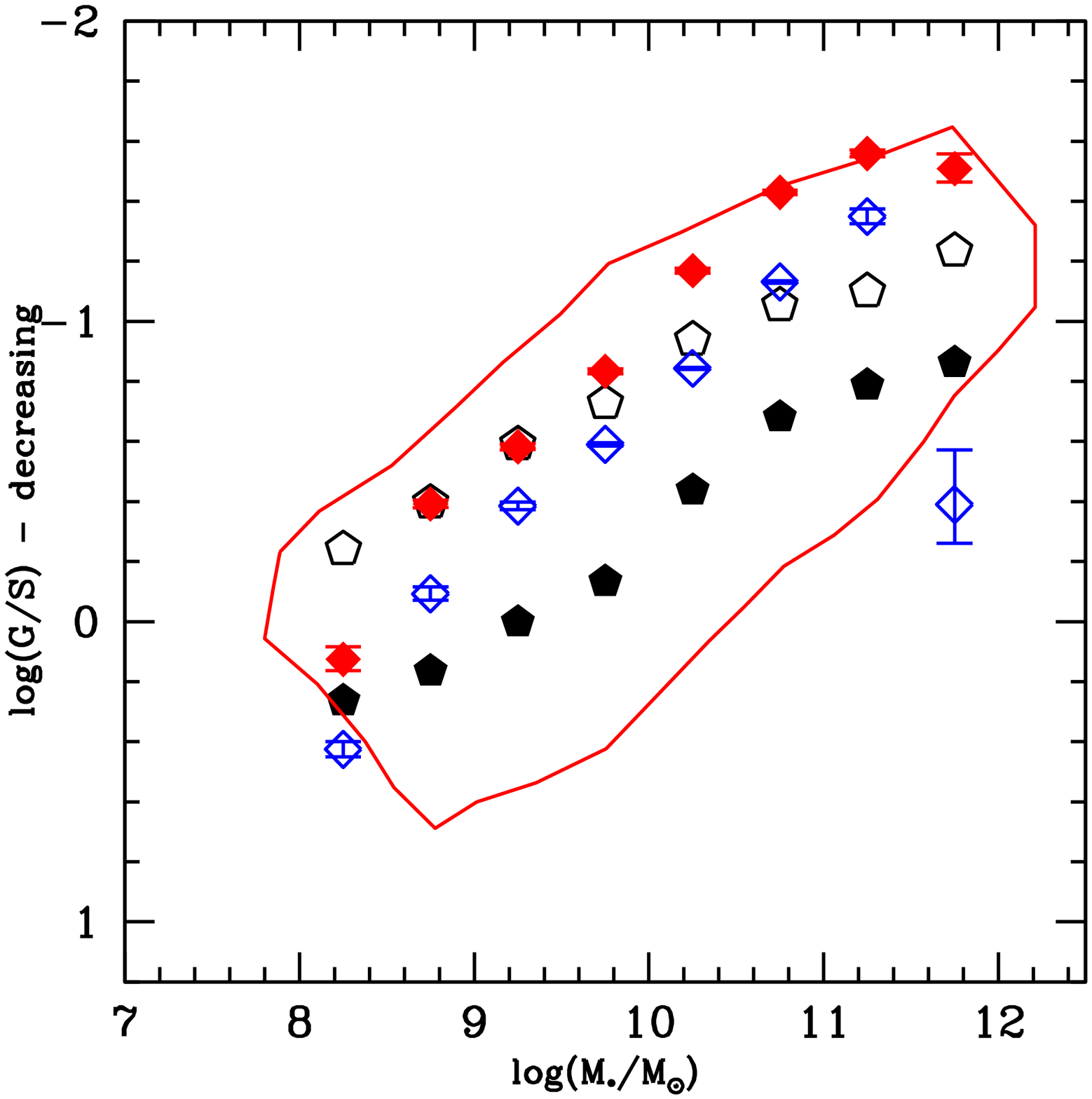}
\caption{G/S  as a  function of the stellar mass for the blue (solid pentagons)
and red (open pentagons) galactic components as computed by Kannapan (2004).
Symbols are as in Fig.~\ref{plotfive}, with the difference that in this case the red and blue components are
the galaxies with the $E(B-V)\le 0.01$ and $E(B-V) > 0.01$, respectively. }
\label{plotsix}
\end{figure*}
All our calculations have been performed by assuming the same scaling relations for all galaxies,
i.e. we have assumed that all our galaxies are self-gravitating disks,
embedded in dark matter halos. This assumption may seem unrealistic, since spheroids are known to follow
scaling relations different from disk galaxies. To tackle this problem, we have divided our galaxy sample into two categories,
the first composed by disk galaxies, the second by spheroids. Disk galaxies follow 
the scaling relations described in section 3.
For spheroids, instead of an exponential surface mass density profile, we have assumed a Sersic law, given by:
\begin{equation}
\Sigma_{sph}(R)= \Sigma'_0 \exp\left\{-b_{n}[(R/R_{eff})^{1/n}-1]\right\}
\label{deV}
\end{equation}
(Sersic 1968), where $b_{n} \simeq 2n-0.324$ (Ciotti \& Bertin 1999). 
We want to test two possible values for 
the index $n$: $n=4$, corresponding to the de  Vaucouleurs (1948) law, and $n=2$, since most galaxies do not fit
into the binary ``de Vaucouleurs'' or ``exponential disk'' categories
but instead span some range of Sersic parameters (Blanton et al. 2003).\\
In this case, for the SFR surface density profile, we have assumed that  $\dot{\Sigma}_{*,sph}(R) \propto \left[\Sigma_{sph}(R)\right]^{1.4}$.
To compute the effective radius $R_{eff}$, we have used the Shen et al. (2003) relation between $R_{eff}$ and $M_{*}$:
\begin{equation}
R_{eff} = 4.16 \left(\frac{M_{*}}{10^{11}M_{\odot}}\right)^{0.56} kpc
\label{spheroid}
\end{equation}
(see also Boylan-Klochin et al. 2005, Robertson et al. 2007), valid at any lookback time.

The distinction between disks and spheroids has been performed by means of two criteria:
(1) the stellar mass of the galaxy and (2) its present day
SFR. The first criterion seems plausible since it is known  that at low redshift, early type galaxies have stellar masses
higher than late types and that the mass function of galaxies at stellar masses $M_{*} < 10^{11}M_{\odot}$ and
$M_{*}\ge 10^{11}M_{\odot}$ is dominated by late types (i.e. disks) and early types
(i.e. possibly spheroids, Pannella et al. 2006, Shankar et al. 2006), respectively. On the basis of this, we have computed the gas masses assuming that
all the galaxies with $M_{*} < 10^{11}M_{\odot}$ obey the scaling relations described in section 3, whereas all the
galaxies with stellar mass $M_{*} \ge 10^{11}M_{\odot}$ are characterized by a Sersic
surface density profile, as given by equation \ref{deV}, and follow the scaling relation of eq. \ref{spheroid}.  
We are aware that this criterion prevents us from adopting the appropriate profile for 
low-mass early type galaxies, such as dwarf spheroidals and dwarf ellipticals. For this reason, we consider 
an alternative criterion for 
the distinction between disks and spheroids, based on the SFR. \\
In the second case, we have assumed that the galaxies with a present-day SFR lower than a given threshold
(we have used a threshold value of $0.5 \, M_{\odot}/yr$)
and having assembled  more than $50 \%$ of their total stellar mass during the first timestep, belong
to the category of spheroids, whereas the rest are spiral disks.
In Fig. ~\ref{plotthree}, we show the present-day gas to stellar mass ratio as a function of the stellar mass,
computed assuming the same scaling laws for all galaxies (solid circles), different scaling laws for spheroids and disks
distinguished by mass (solid squares) and by current SFR. 
While distinguishing by SFR, we plot our results assuming two different values for the Sersic index $n$ of the 
density profile assumed for spheroids: $n=4$ (open circles) and $n=2$ (solid triangles). \\
In  any case, we obtain gas masses and gas fractions very similar to the ones obtained assuming the same scaling relations for
all the galaxy population. The same situation occurs at larger lookback times, i.e. at any redshift, the gas masses
computed in the three cases are very similar. This indicates that our results are fairly independent of the adopted profile. 
Motivated by this fact, all the results presented in the remainder of the paper  have been
obtained by adopting the same disk scaling relations for the whole galaxy sample. \\
At this point, our aim is to understand how our estimates of the scaling radii 
performed by means of the exponential profile compare with the optical radii of the SDSS galaxies, determined from 
the observed light profiles.  In Fig.~\ref{plotfour}, left panel, we show the computed scaling radii as a function of the r-band Petrosian 
radii $R_P$ 
for  the SDSS sample, overplotted with the mean are $\pm$ 1 $\sigma$  errors. In Fig.~\ref{plotfour}, the darkest 
regions correspond to the densest areas, where most of our galaxies lie. 
This figure shows that for the bulk  of the galaxies and 
for radii $\le 8 kpc$, 
there is almost a 1:1 correspondence between scaling radii and Petrosian radii. 
The scaling radii underestimate the optical radii for $R_P \ge 8 kpc$. The reason for this discrepancy may be found in 
Fig.~\ref{plotfour}, 
right panel, where we show the expected angular size, computed  from the scaling radii, against actual angular size on the sky for 
the SDSS galaxies. 
From this plot, we can see that the deviations are largest when the size on sky is very small or very large, and that 
the discrepancy concerns 
the minority of the galaxies, hence it is not a major reason of concern. 
The match between scaling radii and Petrosian radii seem problematic 
when the galaxy radius is $<4$ arcsec and $>18$ arc sec, otherwise there is a very good agreement. 
For a limited number of galaxies, the underestimation of the actual size may imply an underestimation of the gas fractions. \\ 
Galaxies of the SDSS sample have been shown to split into two different populations, one composed of blue
galaxies dominated by recent star formation episodes and another composed of red, passively-evolving
early types (Strateva et al. 2001, Blanton et al. 2003).
In figure~\ref{plotfive}, we show the calculated present day G/S values as a function of the
stellar mass for these two different galactic populations, compared to the
corresponding observational estimates by Kannapan (2004).
The estimates by Kannapan (2004) are based on photometric techniques.
The atomic G/S are estimated using the $u-K$ colours of $\sim 35000$
galaxies from SDSS DR2 and 2MASS databases. This technique is calibrated by means of $HI$ data
from the HyperLeda $HI$ catalogue. By means of her technique, Kannapan (2004) estimated individual
$\log(G/S)$ values between $\sim -2$ and $\sim 0.5$. The G/S estimates are very reliable
in the lowest mass bins but tend to represent overestimations in the largest mass bins
(S. Kannapan, private communication). For this reason, here we compare the average G/S values computed
considering only the galaxies with $\log(G/S)>-2$. \\
By comparing our results with the ones of Kannapan (2004), we are able to obtain constraints on the epoch at which galaxies of the blue
and red populations assembled the bulk of their stars. In particular, we can reproduce very well the data obtained by K04 if we assume that
the red population consists of all the galaxies having formed $>60 \%$ of their
present-day stellar mass during the first 4 timesteps, i.e. in $\sim 11.4$ Gyr. 
The overall agreement between our estimations
and the observational estimates by Kannapan (2004) is good for all galaxies with stellar masses
$\log(M_{*}/M_{\odot}) \le 11$.
For the most massive galaxies of the red population, we tend to underestimate
the gas fractions. This is not surprising, since as stated above, the estimates by Kannapan (2004) are
likely to represent overestimations of the actual values in particular in the highest mass bins.
For the most massive galaxies of the blue population, we predict mean (G/S)
higher than the ones determinded by K04
because of the fact that highest mass bins contain very few galaxies, some of which present very high
gas fractions, considerably increasing the average G/S in these bins.

\begin{figure*}
\centering
\vspace{0.001cm}
\includegraphics[height=15cm,width=15cm]{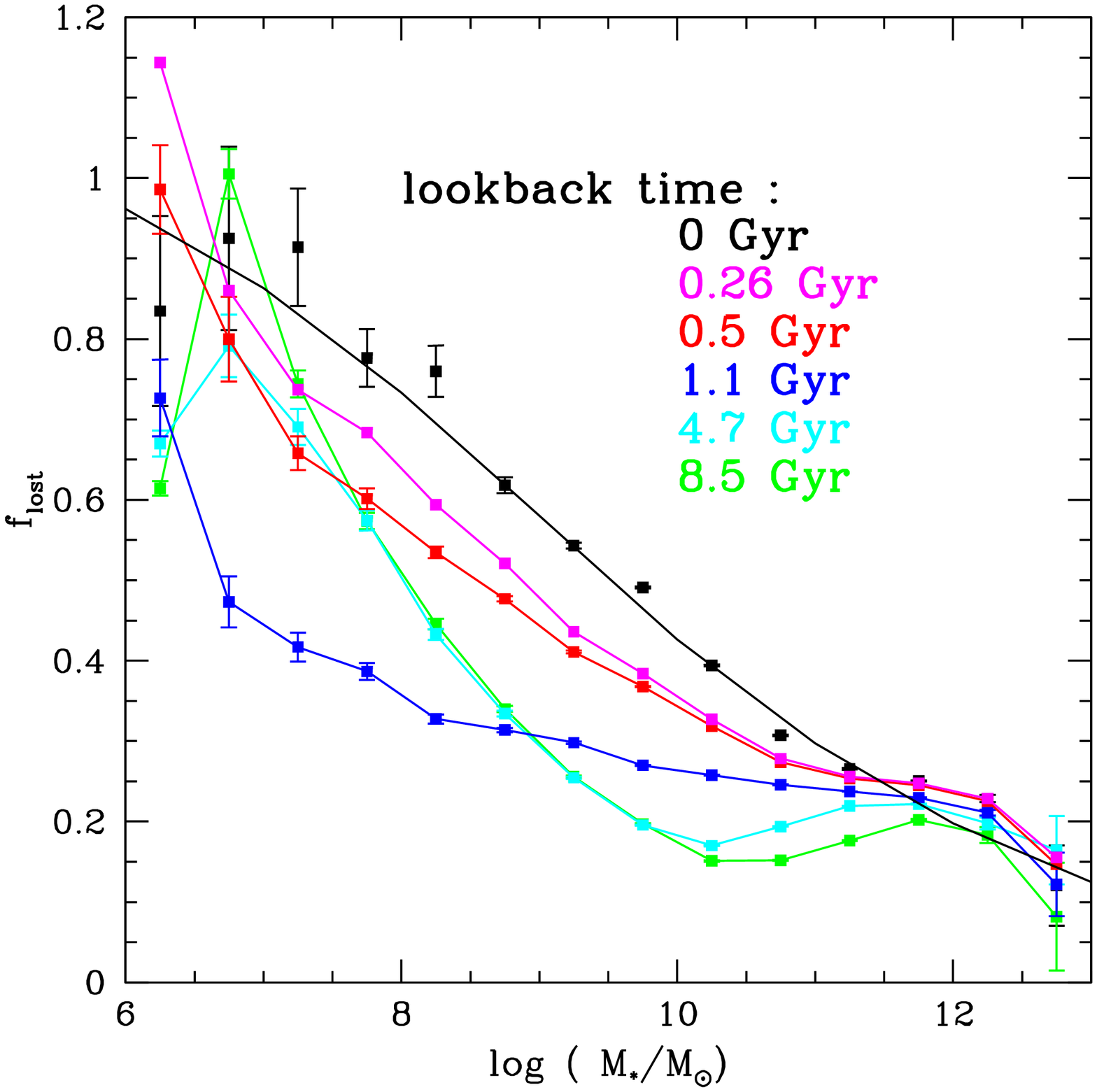}
\caption[]{Time evolution of the mean fraction of lost
baryons $f_{lost}$, as a function of the stellar mass for the
SDSS catalogue.
Various colours correspond to different lookback times.
The green, cyan, blue, red, magenta and black sqaues with error bars are
the $f_{lost}$ as a function of the stellar mass computed at lookback times
of 8.5 Gyr, 4.7 Gyr, 1.1 Gyr, 0.5 Gyr, 0.25 Gyr and 0 Gyr
(i.e. at the present day). The green, cyan, blue, red and magenta lines
are drawn to guide the eye through the points.
The solid black line is an
analytical fit to the present day values (see section 4.2).  The anomalous behaviour at 1.1 Gyr is almost certainly a result of the difficulty of estimating the star formation rate at this age (see text for more details).}
\label{plotseven}
\end{figure*}
\begin{figure*}
\centering
\vspace{0.001cm}
\includegraphics[height=15cm,width=15cm]{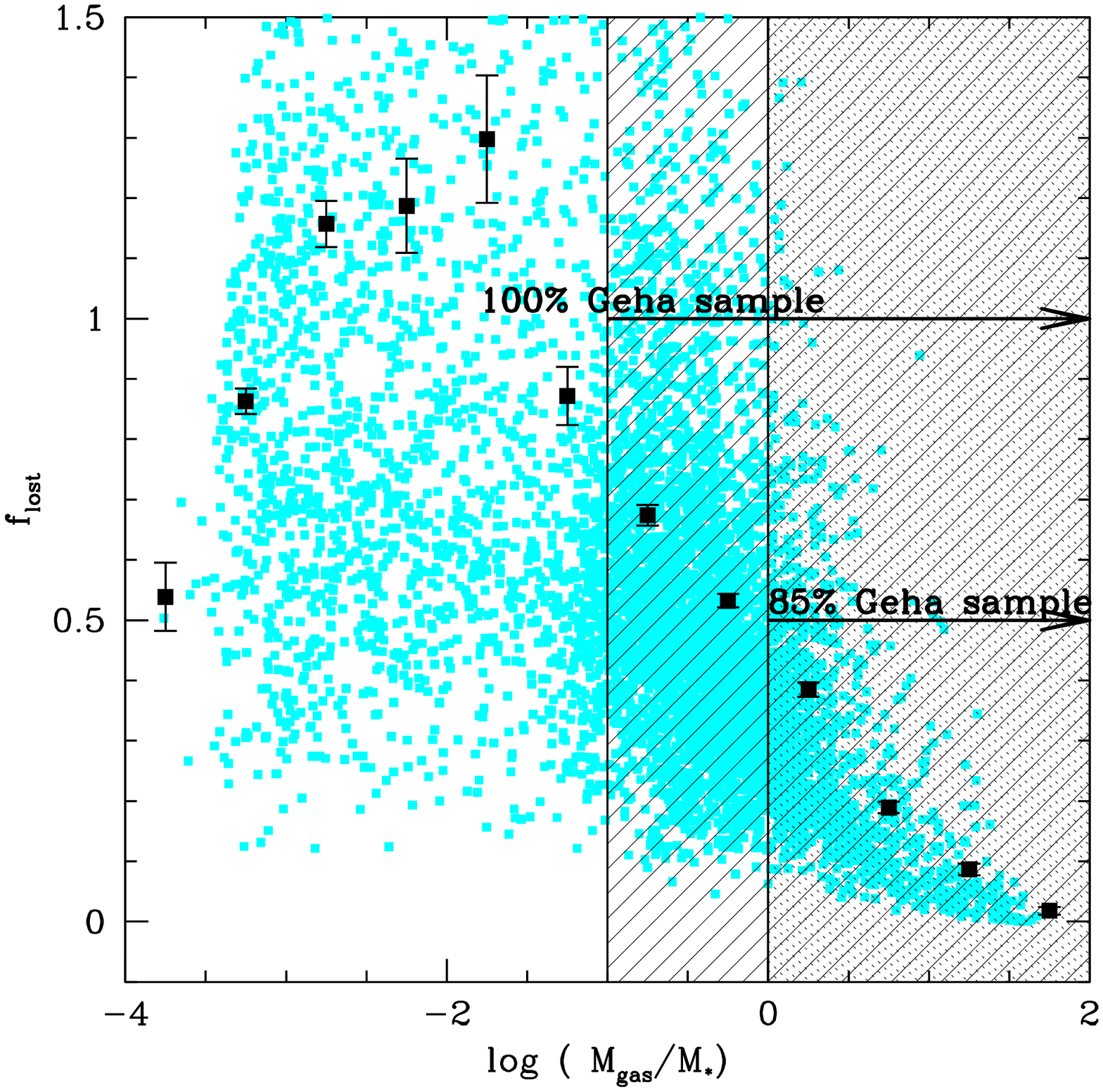}
\caption[]{$f_{lost}$ vs G/S for all the
dwarf galaxies of our sample, i.e. all the galaxies with $M_{*} \le 10^{9} M_{\odot}$. 
The small solid squares are the points for 
all the individial galaxies, whereas the large squares are the mean values, 
plotted with the error bars. 
The larger shaded area  and the smaller dotted area indicate the 
regions where 100\% and 85 \% of the dwarf galaxies of the sample 
by Geha et al. (2006) lie, respectively. }
\label{ploteight}
\end{figure*}

The study by Panter et al. (2007b, in preparation)
shows that it is possible to split the SDSS galaxies into two populations on the basis of their
colour excess $E(B-V)$,  which expresses the amount of dust present in any galaxy.
To disentangle between the blue and the red galactic populations, we tested also this criterion.
In this case, the red and blue components are
the galaxies with the $E(B-V)\le 0.01$ and $E(B-V) > 0.01$, respectively.
In figure~\ref{plotsix}, we present the predicted mean (G/S) values for the red and blue galaxy populations as compared
to the observational data by K04, having determined the distinction between the two populations by means
of their colour excess $E(B-V)$.
The separation between blue and red appears less clear than in the former case.
Furthermore, the agreement between our predictions and the data by K04 is worse than the one shown in Fig.~\ref{plotfive}.
In fact, in the case of Fig.~\ref{plotsix}, the observed (G/S) are systematically
underestimated by our predictions, in particular the data for the blue galaxy population.
This shows that the colour excess is not a totally reliable diagnostic to disentangle blue and red galaxies,
being  strongly dependent on the inclination causing overlapping between the two different populations (Calzetti 2001).\\
To conclude this section, we compare our results for the estimated gas fractions with some individual cases of the sample 
studied by Geha et al. (2006).\\
G06 studied the gas content for a sample of low luminosity dwarf galaxies selected from the Sloan Digital Sky Survey.
This sample consits of 101 galaxies, for which they  have obtained follow-up $H_{I}$ 
observations using the Arecibo Observatory and Green Bank Telescope. Using the measured masses 
of $H_{I}$ and the optical properties, G06  derived the $H_{I}$-to-stellar mass ratio   $M_{H_{I}}/M_{*}$ 
for the galaxies of their sample.\\
We have indentified 23 galaxies from our sample that are also present in the G06 sample. 
For this 23 galaxies, we compare our values for the stellar masses and $M_{H_{I}}/M_{*}$ to the values found by Geha et al. (2006) 
by means of a different technique. The results of this comparison are shown in table 1.  
For each galaxy, we derive the quantities $M_{H_{I}}$  by multiplying the gas mass  $M_{g}$ by the solar photospheric 
H mass fraction  $X_ {H}=0.75$ (Lodders 2003). \\
In some cases, the stellar mass values estimated by means of the MOPED algorithm differ from the values reported by G06. 
These differences may be due to several factors, such as the assumption of a different IMF 
(G06 use the IMF by Kroupa et al. 1993) and 
uncertainties in the distances because of peculiar velocities. G06 do not present any uncertainty for their determined 
stellar masses. 
For 11 out of 24 galaxies, the discrepancies between our $M_{H_{I}}/M_{*}$ values and the ones found by G06 are within 
0.7 dex, corresponding to a factor of 5. In 10 cases, we severely underestimate the $M_{H_{I}}/M_{*}$ and in the few remaining 
cases, we overestimate the results by G06 by a factor larger than 5. 
The very low $M_{H_{I}}/M_{*}$ we obtain in some cases may be due to underestimated 
galaxy dimensions or to the presence of infalling gas not taking part to star formation, as observed for the local Sculptor dwarf 
spheroidal galaxy (Carignan et al. 1998). However, most likely it is an indication of the fact that MOPED cannot determine precisely the star formation history of individual galaxies but of ensambles of them, and only for large numbers the errors are very small (see also Fig.~8). There is however another trend worth pointing out: those galaxies for which we find the best agreement in the Geha sample are the ones that show strongest absorption lines, thus providing MOPED a better handle to recover the star formation history, while those that give the worse agreement do have only emission lines (recall that we only model the continuum and absorption lines with MOPED.

\renewcommand{\baselinestretch}{1.0}
\begin{table*}
\centering
\caption{$M_{H_{I}}/M_{*}$ calculated for single galaxies and 
compared to the results by Geha et al. (2006). }
\begin{tabular}{lcccccccc}
\\[-2.0ex] 
\hline
\\[-2.5ex]
\multicolumn{1}{l}{Name}&\multicolumn{2}{c}{This paper}&\multicolumn{1}{c}{}&\multicolumn{2}{c}{Geha et al. (2006)}&\multicolumn{1}{c}{}&\multicolumn{2}{c}{}\\
\hline
\multicolumn{1}{c}{}&\multicolumn{1}{c}{log($M_*$)}&\multicolumn{1}{c}{log($M_{H_{I}}/M_{*}$)}&\multicolumn{1}{c}{}&\multicolumn{1}{c}{log($M_*$})&\multicolumn{1}{c}{log($M_{H_{I}}/M_{*}$)}&\multicolumn{1}{c}{$\sigma_{HI}$}\\
\hline
\hline
\\[-1.0ex]
386141    &    6.95583   &   0.47055  &    &     7.82000   &  -0.26000   &  0.08     &  \\   
192963    &    7.83769   &   0.11271  &    &     7.96000   &   0.82000   &  0.04     &  \\   
192971    &    8.05618   &  -0.81446  &    &     8.00000   &   0.52000   &  0.05     &  \\   
190632    &    8.28466   &  -2.06788  &    &     8.32000   &   0.46000   &  0.03     &  \\   
677002    &    8.06461   &  -2.74395  &    &     7.85000   &   0.33000   &  0.08     &   \\  
 47936    &    8.41572   &  -0.59873  &    &     8.51000   &   0.26000   &  0.06     &   \\  
222989    &    8.08221   &  -0.00197  &    &     8.10000   &   0.63000   &  0.06     &   \\  
203478    &    7.88343   &  -0.07654  &    &     8.17000   &   0.40000   &  0.03     &   \\  
191112    &    7.44543   &  -0.89164  &    &     7.66000   &   0.38000   &  0.07     &   \\  
 48406    &    8.22616   &  -0.58562  &    &     8.30000   &  -0.32000   &  0.12     &   \\  
231588    &    7.62681   &   0.12001  &    &     7.59000   &   0.67000   &  0.07     &   \\  
227294    &    7.36842   &  -1.51760  &    &     7.77000   &   0.11000   &  0.07     &   \\  
462731    &    6.85259   &   0.89671  &    &     7.55000   &   0.26000   &  0.08     &   \\  
467776    &    8.15761   &  -0.59137  &    &     8.09000   &   0.10000   &  0.09     &   \\  
123408    &    7.92148   &  -0.22780  &    &     8.09000   &  -0.28000   &  0.08     &   \\  
132909    &    7.32389   &   0.11341  &    &     7.61000   &   0.65000   &  0.09     &   \\  
136373    &    7.35887   &  -2.07113  &    &     7.84000   &  -0.31000   &  0.09     &   \\  
276603    &    7.92763   &  -1.07278  &    &     7.72000   &   0.53000   &  0.07     &   \\  
232890    &    7.59235   &   0.02063  &    &     7.78000   &   0.42000   &  0.07     &   \\  
169071    &    6.63464   &   0.06289  &    &     7.43000   &  -0.38000   &  0.20     &   \\ 
278622    &    7.53275   &   0.33543  &    &     8.07000   &   0.06000   &  0.08     &  \\  
262647    &    6.67881   &   1.05205  &    &     7.86000   &  -0.30000   &  0.14     &  \\   
565755    &    7.39599   &  -3.14264  &    &     7.96000   &   0.29000   &  0.11     &  \\       
\hline
\hline
\end{tabular}
\flushleft
\end{table*}

\subsection{Fractions of lost baryons in SDSS galaxies}
\label{ouflow}
In galaxies, gas can be both transformed into stellar form 
and ejected into the external environment, i.e.  the intergalactic or intracluster media (ICM, IGM).  
Gas can be ejected by galaxies by means of various processes. 
Supernova (SN)  explosions heat  
the interstellar gas and, 
as soon as the binding energy of the gas 
exceeds its thermal energy, 
some fraction of the ISM can be removed through SN-driven galactic winds (e.g., Larson 1974). 
In general, isolated galaxies eject gas mostly by means of galactic winds. 
Galaxies in dense environments (i.e. clusters and groups) undergo gas loss via additional mechanisms, 
which in general  
are bound to environmental effects 
and can be of various types: tidal interactions, ram pressure stripping, viscous stripping, starvation and thermal evaporation 
(see Boselli \& Gavazzi 2006 and references therein). 
In this section, we aim at studying how the total amount of cold gas ejected by means of any mechanism 
evolved with cosmic time. Our approach does not allow us to infer which are the 
main mechanisms driving gas ejection, for which more complex dynamical simulations 
or galactic chemical evolution models would be suited, where SN feedback is taken into account (Matteucci 1994; Recchi et al. 2002). \\
To perform this task, once again we use the 11 timebins described in sect.~\ref{MOPED}.
For each galaxy, at each timebin  $t_{i}$ we know the total baryonic 
mass $M_{tot}(t_{i})=M_{gas}(t_{i})+M_{*}(t_{i})$.
We can compute the net ejected (or accreted) mass as the 
difference between the total baryonic mass computed at two following timesteps $t_{i-1}$ and $t_{i}$,
$M_{ej} = M_{tot}(t_{i}) - M_{tot}(t_{i-1})$. This quantity may be positive or negative, with the negative results indicating mass loss.  
Note that gas which is heated to the extent that it does not take part in star formation is counted as ejected, 
regardless of whether it is physically removed from the galaxy or not. 
We can use this information to compute the complete mass outflow histories for the galaxies of our sample.
When baryons are lost or accreted, the total mass of the galaxy goes down,
which puts it in a different mass bin. 
The current stellar mass of a galaxy depends on the star formation hostory of that galaxy, which is used here to 
determine the infall and ouflow history.\\ 
In Figure~\ref{plotseven}, we show the computed evolution of the mean fraction of lost baryons $f_{lost}$, 
defined as $f_{lost}=\frac{M_{ej}}{M_{*}+M_{gas}}$, as a function
of the stellar mass for the SDSS catalogue.
With one exception, these curves show a very consistent picture of progressive baryon loss over time for all but the most massive galaxies, with the lowest-mass systems losing much more baryon mass than the high-mass systems.  The exception to this is the curve at a lookback time of 1.1 Gyr, which is anomalous.  The reason for this is almost certainly connected with the difficulty of determining the star formation rate for this population, as there is nowhere in the spectrum where the population is dominant (Mathis et al, 2006). 

The mean $f_{lost}$ values computed for large galaxies ($\log(M_{*}/M_{\odot})>11$) do not show a strong evolution with time, remaining at a low level of $\sim 10-20\%$.  Combined with the G/S ratio for these systems, we conclude that early star formation is very efficient in these galaxies. 
On the other hand, the mean $f_{lost}$ values computed for galaxies in the range  $7\le \log(M_{*}/M_{\odot}) < 11$,
undergo a strong evolution with lookback time, and this indicates that
the galaxies of lowest mass are continually losing baryonic mass from the cold phase over a long period. 
This result is a confirmation of the predictions
from chemical evolution models for elliptical galaxies (Matteucci 1994),
which showed that, in order to explain the increase of the [Mg/Fe] ratio  with galactic mass,
one has to assume that the efficiency of star formation is an increasing function of the galactic mass.
One possibility for 
this is an ``Inverse Wind'' scenario of galaxy evolution, in which the most massive galaxies are the first ones
to experience the galactic winds and to complete their outflow and star formation history, whereas the dwarf galaxies undergo substantial outflows
and continuous star formation until most recent times.

As one would expect from the depths of the potential wells, we find very clear evidence that at the present day, the lower-mass galaxies have lost a larger fraction of their cold baryons, as much as 90\% for galaxies with a stellar mass of $10^6 M_\odot$.
An analytical fit to the present day $f_{lost}$-$M_{*}$ relation is
\begin{equation}
f_{lost}=0.58 - 0.51 \,\, {\rm atan} \left\{0.31\left[\log\left(\frac{M_{*}}{M_\odot}\right)-9.0\right]\right\},
\end{equation}
valid for $6<\log(M_*/M_\odot)<12.5$.
An increasing trend of the ejected fractions with decreasing galactic mass is also confirmed by analytical and numerical single-burst  models for
dwarf galaxies (Mac Low \& Ferrara 1999). Our results are in good agreement
with results from galactic chemical and chemo-dynamical models considering more complex star formation histories.
By means of chemo-dynamical simulations, Recchi et al. (2002) showed that a dwarf galaxy of baryonic mass of $\sim 10^{7} M_{\odot}$
undergoing
multiple starbursts can eject up to $\sim 75 \%$ of its mass.  Chemical evolution models for elliptical galaxies of masses between
$10^{9} M_{\odot}$ and $10^{12} M_{\odot}$ predict decreasing ejected fractions,
with values  between $90 \%$ and 10 $\%$ (Gibson \& Matteucci 1997, Pipino et al. 2002).

The shape and normalization of the present day $f_{lost}$ vs $M_{*}$ curve derived
observationally in this study is in good agreement with theoretical
predictions from chemo-dynamical models where only supernova feedback
has been employed. While we cannot rule out that AGN feedback can be
also at play with the present observations, it seems that supernova
feedback alone is able to displace the observed amount of cold gas
from the dark halo as a function of mass.\\
This study does not exclude preheating (Mo et al. 2005, Crain et al. 2007), which prevents accretion
of gas, but rather provides evidence of some process which has ejected gas
from the cold phase after accretion. 

Finally, in fig.~\ref{ploteight}, we show the $f_{lost}$ vs G/S for all the
dwarf galaxies of our sample, 
in this case all the galaxies with $M_{*} \le 10^{9} M_{\odot}$. 
Here, we show the 
behaviour of all the individial galaxies and the mean values, plotted with the error bars. 
Fig.~\ref{ploteight} shows a clear anticorrelation between the lost fraction and the present-day gas fraction.  
The two average values computed in the lowest G/S bins apparently 
show an anomalous behaviour, but these two bins contain very few systems, 8 total. 
The larger shaded area  and the smaller dotted area indicate the 
regions where 100\% and 85 \% of the dwarf galaxies of the sample 
by G06 lie, respectively. 
Fig.~\ref{ploteight} shows that our results are in substantial agreement with the ones 
of G06.
According to Geha et al. (2006), the baryonic Tully-Fisher relation 
indicates that gas removal processes 
were not important for the majority of their sample.  
The Geha sample includes dwarf galaxies with extremely high gas fraction, which, according to our results, 
must have retained the majority ($> 60\%$ if we consider the $85\%$ of the Geha galaxies) of their present mass.\\
Our study strongly indicates that
the main driver of the fraction of lost gas is the mass of the galaxy.
Environmental effects  could also play a role in determining the 
fraction of  gas lost from galaxies, in particular for low mass galaxies
(Geha et al. 2006). 
However, the SDSS galaxies span a large range of different environments
and if the environmental effects were dominant, we should not expect
such a clear correlation
between lost gas and stellar mass. Furthermore,  
recent studies  (Sheth et al. 2006, Blanton et al. 2006, Mateus et al. 2007, Berta et al. 2008) 
of the relation between galaxy colours, luminosities and
star formation histories  and environment
for the SDSS sample have shown that such properties are more related to
the mass of the host dark matter halo than to the large scale
environment. The marked correlation studies in Mateus et al. (2007) and Berta et al. (2008) show that 
the environmental dependence is mostly driven by the mass, although other physical parameters like spin do play a secondary role.
A more detailed study of lost gas fraction as a function of environment
is deferred to a future paper. \\
The results of this paper have been obtained self-consistently from the star formation history of the SDSS galaxies,
in a way which should be robust to the degree of merging which has taken place.
The star formation history is linked, via the
Kennicutt law, to the gas mass present at any time, and hence allows us to trace the net gas accretion/outflow history. 
The metallicity history can be used to determine the metal outflow history, the lost metal fractions
and the chemical enrichment of the inter galactic medium; this will the subject of a forthcoming paper.

\section{Conclusions}
In this paper, we used the Kennicutt (1998) relation, linking the surface
star formation rate to the gas mass surface density,
together with the scaling relations by Mo, Mao \& White (1998), linking the
galactic scale-length radius to the baryonic
mass, to recover the gas masses for
$\sim 310000$ galaxies of the SDSS DR3 sample. Our main conclusions can be summarised as follows. \\
1) We studied the time evolution of the gas to stellar mass fractions for all  the galaxies of the SDSS catalogue.
In the last Gyr of evolution, our predictions indicate an increase in
the average (G/S) for the galaxies with stellar masses $M_{*}<10^{10} M_{\odot}$, in agreement with the
the downsizing picture of galaxy evolution, where
the most massive galaxies form the bulk of their stars at early times.
At the present time, we predict individual
$\log(G/S)$ values between $-4$ and $\sim 1.5$. These lower and upper limits are
 in agreement with independent estimates of the (G/S) in the local gas poorest and richest galaxies, respectively. \\
2) We split the galaxy sample into a red and a blue population  by means of two different criteria,
i.e. the recent amount of star formation and the colour excess $E(B-V)$.
The (G/S) calculated by adopting the first criterion are in good agreement with independent
estimates based on photometric techniques (Kannappan 2004),
in particular for the galaxies with stellar masses
$\log(M_{*}/M_{\odot}) \le 11$. The adoption of  the colour excess criterion implies instead a less clear distinction between the
two galactic populations. \\
3) We computed the time evolution of the average fraction of lost baryons $f_{lost}$, defined as $f_{lost}=\frac{M_{ej}}{M_{*}+M_{gas}}$
as a function of the stellar mass for the SDSS catalogue.
With the exception of an anomaly at a lookback time of 1Gyr (which we put down to the difficulty of determining star formation rates at this time; Mathis et al 2006), the results show clear signatures that the fraction of lost cold baryons increases with time for all but the highest-mass galaxies, and that the fraction lost is heavily dependent on the stellar mass of the galaxy.  The lost fraction varies between about 10-20\% for $M_*>10^{12}M_\odot$, to $\sim 80-90\%$ for low-mass ($M_{*}\le 10^{7} M_{\odot}$) galaxies. 
The significant loss of gas from the low-mass systems is not surprising given the small potential wells involved, and may be responsible for the lack of observed low-mass galaxy satellites compared with simulations (Klypin et al 1999, Moore et al 1999), even including the latest dwarfs detected in SDSS (e.g. Belokurov et al 2007).
Our results are in agreement with chemical evolution models for elliptical galaxies (Matteucci 1994), and could be explained by
an ``Inverse Wind'' scenario for galaxy evolution, i.e. that the most massive systems are the first ones to undergo mass loss
via galactic winds. In any event our results show clearly the ``downsizing'' character of the galaxy population, with the star formation in high-mass galaxies essentially complete at early times. 

\section*{Acknowledgments}
We wish to thank Simone Recchi, Sheila Kannappan, Daniela Calzetti and Antonio Pipino for many interesting
discussions. FC thanks the hospitality of the Department of Physics \& Astronomy of the University of Pennsylvania,
where part of this work was carried out. We thank an anonymous referee for comments and suggestion that helped us improve the paper.

 \acknowledgments

 \end{document}